\documentclass[prl, showpacs, twocolumn, floatfix]{revtex4}

\usepackage{graphicx}
\usepackage{amsmath, amsfonts, amssymb, bm}
\usepackage[]{psfrag}

\psfragscanoff
\begin{document}

\title{ Two-center resonant photoionization  }

\author{B.Najjari}
\author{A.B.Voitkiv}
\author{C.M{\"u}ller}
\affiliation{Max-Planck-Institut f\"ur Kernphysik, Saupfercheckweg 1, 69117 Heidelberg, Germany}

\date{\today}

\begin{abstract}
Photoionization of an atom $A$, in the presence of a neighboring atom $B$, 
can proceed via resonant excitation of $B$ 
with subsequent energy transfer to $A$ through 
two-center electron-electron correlation. 
We demonstrate that this two-center mechanism can 
strongly outperform direct photoionization at nanometer 
internuclear distances and possesses characteristic features 
in its time development and the spectrum of emitted electrons. 
\end{abstract}
 
\pacs{
32.80.-t, %(Photoionization and excitation)
32.80.Hd, %(Auger effect)
%82.30.Cf, %(Atom and radical reactions; chain reactions; molecule-molecule reactions)
33.60.+q, %(Photoelectron spectra)
82.50.Hp %(Photochemistry: Processes caused by visible and UV light)
} 

\maketitle

Ever since Einstein proposed his interpretation 
of the photoelectric effect \cite{Einstein}, photoionization (PI) 
studies on atoms and molecules have played a key role for our 
understanding of the basic laws of quantum physics. 
Modern PI experiments providing complete information on all quantum degrees 
of freedom allow for stringent tests of the most advanced calculations \cite{Martin}. 
A new era of PI studies is presently being opened by the worldwide emergence 
of advanced light sources such as x-ray free-electron laser (XFEL) 
facilities (see \cite{XFEL3} and references therein). % \cite{XFEL1,XFEL2,XFEL3}. 

The structure and time evolution of matter on a microscopic scale crucially depends on
electron-electron correlations. Their influence ranges from atoms 
and small molecules to organic macromolecules and solids. 
Electron correlations are responsible for deexcitation  reactions 
in slow atomic collisions \cite{Smirnov} 
and quantum gases \cite{Weidemueller}. They play a prominent 
role in energy transfer between chromophores \cite{Forster} 
and lattice dynamics in polymers \cite{Suhai}. 
They also represent the origin of magnetism and superconductivity \cite{solids}. 
Another effect caused by electron correlations 
is ultrafast intermolecular decay of inner-valence vacancies 
which has been recently observed in various rare gas dimers 
and clusters \cite{clusters,resonantICD,He} 
and water molecules \cite{ICDexpH2O}.
Also electron-ion recombination 
can be greatly enhanced by the presence of 
a neighboring atom \cite{2CDR}. 

PI may reveal particularly clean manifestations 
of electronic correlations. Prominent examples are 
single-photon double ionization \cite{SPDI},
laser-induced autoionization \cite{Rzazewski},
and non-sequential double ionization in strong laser fields \cite{NSDI}.

Against this background we study in this Letter photo ionization, 
which involves resonant electronic correlations between  
two neighboring atomic centers (atoms, ions or molecules). 
In this process, which may be termed two-center photo ionization (2CPI), 
one of the reaction pathways for ionization of an atom is  
radiationless transfer of excitation from a neighboring atom,  
whose bound states are resonantly coupled by 
the external electromagnetic (EM) field (see Fig.1). 
Characteristic properties of 2CPI 
are revealed both in quite weak and more intense EM fields. 
In case of weak fields PI can be enhanced by orders 
of magnitude in the presence of a neighboring atom. 
The case of more intense fields is characterized 
by a step-wise development of the ionization in time and 
multiple peaks in the energy spectrum of photoelectrons. 
Corresponding experiments may be feasible at synchrotron or XFEL beamlines. 
Our study thus connects the currently very 
active research areas of interatomic phenomena 
\cite{clusters,resonantICD,ICDexpH2O,He,2CDR} and field-induced  
PI dynamics \cite{XFEL3}. %\cite{XFEL1,XFEL2,XFEL3}. 
Another example of this topical combination 
are decay mechanisms 
in % doubly-excited \cite{CederbaumNew} or 
multiply ionized clusters \cite{Rost} after XFEL irradiation.

%(The metastable initial state in ICD is typically prepared by PI (or %photoexcitation \cite{Ne2}). 
%In the case of He dimers, one of the He %atoms first was simultaneously ionized 
%and excited by photo-absorption from a synchrotron beam.)

%Atomic units (a.u.) are used throughout unless otherwise stated.

\begin{figure}[b]  
\vspace{-0.25cm}
\begin{center}
\includegraphics[width=0.27\textwidth]{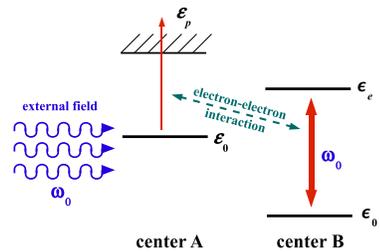}
\end{center}
\vspace{-0.5cm} 
\caption{ \footnotesize{ 
Scheme of two-center photo ionization (2CPI). }} 
\label{figure1}
\end{figure}

In order to understand the basics of 2CPI, 
we consider PI in a very simple atomic system  
consisting of two one-electron atoms 
($A$ and $B$). Both are initially in their ground states 
and separated by a distance $R$ large enough, 
such that one can still speak about individual atoms.   
Let, for definiteness, the ionization potential $I_A = - \varepsilon_0$  
of the atom $A$ be smaller than 
the excitation energy $\Delta E_B = \epsilon_e - \epsilon_0 $ 
of a dipole allowed $1s$-$np$ transition 
in the atom $B$. If such a system is 
irradiated by an EM field with   
frequency $\omega_0 \approx \Delta E_B$, 
the presence of the atom $B$ may have 
a substantial influence on the ionization process.  
Indeed, in such a case the atom $A$ can be ionized 
not only directly but also via resonance photoexcitation 
of the atom $B$ into the $nl$-state with its consequent deexcitation 
through the transfer of energy $\Delta E_B$ 
to the atom $A$ which results in ionization of the latter.    
Obviously 2CPI cannot occur in homoatomic systems. 

Let us suppose that the atomic nuclei having charge numbers  
$Z_A$ and $Z_B$, respectively, 
are at rest. We take the position of the nucleus $Z_A$ 
as the origin and denote the coordinates 
of the nucleus $Z_B$, the electron of the atom $A$ 
and that of the atom $B$ by ${\bf R}$, ${\bf r}_1$ 
and ${\bf r}_2 ={\bf R} + \mbox{\boldmath{ $\xi$ }} $, 
respectively, where $\mbox{\boldmath{ $\xi$ }}$ 
is the position of the electron of the atom $B$ 
with respect to the nucleus $Z_B$. 

Our consideration of the photoionization process 
is based on the equation \cite{at-units} 
%\begin{eqnarray} 
%i \frac { \partial \left| \Psi \right\rangle }{\partial t }= 
%\left( \hat{H}_0  + \hat{W} + \hat{V}_{rad} \right) \left| \Psi \right\rangle.   
%\label{schroedinger} 
%\end{eqnarray} 
\begin{eqnarray} 
i \partial_t \left| \Psi \right\rangle = 
( \hat{H}_0  + \hat{W} + \hat{V}_{rad} ) \left| \Psi \right\rangle.   
\label{schroedinger} 
\end{eqnarray} 
Here $\left| \Psi \right\rangle$ is the state vector 
of the system consisting 
of the atoms $A$, $B$ and the radiation field, 
$ \hat{H}_0 $ is the sum of the Hamiltonians 
for the noninteracting atoms $A$ and $B$ and the free radiation field,  
$\hat{V}_{rad}$ the interaction of the electrons 
with the radiation field  
and $\hat{W}$ is the interaction of the electrons 
with the external EM field. The latter will be  
taken as a classical, linearly polarized field,  
described by the vector potential 
${\bf A}= c {\bf F}_0/\omega_0 \cos\left(\omega_0 t - {\bf k}_0 \cdot {\bf r}\right)$,
where $\omega_0 = c k_0 $ and ${\bf k}_0$ 
are the angular frequency and wave vector, 
$c$ is the speed of light    
and ${\bf F}_0$ is the field strength. 
The interaction $\hat{W}$ then reads   
\begin{eqnarray} 
\hat{W} &=& \hat{W}_0^+ \exp(-i \omega_0 t) + \hat{W}_0^- \exp( i \omega_0 t) 
\nonumber \\ 
\hat{W}_0^{\pm} &=& \sum_{j=1,2} \exp\left( \pm i {\bf k}_0 \cdot {\bf r}_j \right)    
\frac{{\bf F}_0}{ 2 \omega_0} \cdot \hat{\bf p}_j,   
\label{interaction} 
\end{eqnarray} 
where $\hat{\bf p}_j$ is the momentum operator for 
the $j$-th electron. 

In order to treat the interaction of the electrons 
with the radiation field we adopt the covariant approach 
in which the radiation field is described by 
four potentials in a Lorentz gauge and 
the interaction between the electrons 
is mediated by the exchange of so-called 
transverse, longitudinal and scalar photons 
(see e.g. \cite{mand-shaw}). Note that within such 
an approach the electron-electron interaction is fully 
determined by the coupling of the electrons to the radiation field. 

In the process under consideration 
we have essentially four different basic 
two-electron configurations: (I) 
$\psi_g = u_0({\bf r}_1) \chi_0(\mbox{\boldmath{ $\xi$ }}) $ -- both electrons are in 
the corresponding ground states $ u_0 $ and $ \chi_0 $; 
(II) $\psi_a = u_0({\bf r}_1) \chi_e(\mbox{\boldmath{ $\xi$ }})$, 
in which the electron of the atom $A$ 
is in the ground state while the electron of the atom 
$B$ is in the excited state $\chi_e$; 
(III) $\psi_{{\bf p},g} = u_{\bf p}({\bf r}_1) \chi_0(\mbox{\boldmath{ $\xi$ }})$ 
-- the electron of the atom $A$ is in a continuum state $u_{\bf p}$ 
and the electron of the atom $B$ in the ground state; and 
(IV) $\psi_{{\bf p},e} = u_{\bf p}({\bf r}_1) \chi_e(\mbox{\boldmath{ $\xi$ }})$ -- 
the electron of the atom $A$ is in a state $u_{\bf p}$ 
while the electron of the atom $B$ is 
in the state $\chi_e$. 

The radiation field is initially in its vacuum state $\left| 0 \right\rangle$ 
and then undergoes a transition into a state 
$\left| {\bf k}, \lambda  \right\rangle$ in which one 
transverse ($\lambda=1,2$), longitudinal ($\lambda=3$) 
or scalar ($\lambda=0$) photon with momentum ${\bf k}$ 
is present. 

Taking all this into account, 
one can look for the solution for $ \left| \Psi \right\rangle $ 
by expanding it into the 'complete' set of quantum 
states according to  
\begin{eqnarray} 
\left| \Psi \right\rangle &=& \left( g \psi_g + a \psi_a 
+ \int d^3 {\bf p} \alpha_{\bf p} \psi_{{\bf p},g} \right. 
\nonumber \\ 
&& \left. 
+ \int d^3 {\bf p} \beta_{\bf p} \psi_{{\bf p},e} \right) \left| 0 \right\rangle  
%\nonumber \\ 
%&& 
+ \sum_{ {\bf k}, \lambda } 
\left( g_{{\bf k}, \lambda} \psi_g    
+ a_{ {\bf k}, \lambda } \psi_a  \right.  
\nonumber \\ 
&&  
\left. + 
\int d^3 {\bf p} \alpha^{ {\bf k}, \lambda }_{\bf p} \psi_{{\bf p},g}  + 
\int d^3 {\bf p} \beta^{{\bf k}, \lambda }_{\bf p} 
\psi_{{\bf p},e} \right) \left| {\bf k}, \lambda \right\rangle.  
\label{state_vector} 
\end{eqnarray} 
By inserting (\ref{state_vector}) into (\ref{schroedinger}) 
one obtains a set of differential equations for the unknown 
time-dependent coefficients $g$, $a$, $\{ \alpha_{\bf p} \}$, 
$\{ \beta_{\bf p} \}$, $\{ g_{{\bf k}, \lambda} \}$,     
$\{ a_{{\bf k}, \lambda } \}$, 
$\{ \alpha^{ {\bf k}, \lambda }_{\bf p} \}$ 
and $\{ \beta^{{\bf k}, \lambda }_{\bf p} \}$. 
These equations can be solved analytically 
if one uses the first order perturbation %({\it 1B}) 
theory or the rotating-wave approximation % ({\it RWA}) 
with respect to the interaction $\hat{W}$. %between the electrons and the external EM field. 

Solving these equations also yields 
the (effective) interaction $\hat{V}^{ee}$ 
between the electrons. Although the motion 
of the electrons is nonrelativistic, 
this interaction nevertheless has, in general, 
to account for the retardation effect. 
It becomes of great importance when the time $ \tau_{pr} \sim R / c $, 
which the light needs for traversing the distance between the electrons, 
compares with or even exceeds the effective time $\tau_e \sim 1/\omega_0$ 
of the electron transitions. For electrons undergoing  
electric dipole transitions the interaction $V^{ee}$ reads 
\begin{eqnarray} 
\hat{V}^{ee} &=& r_{1 i} \, \xi_j \, \Theta_{ij}  
\label{AB_inter} 
\end{eqnarray} 
where $ r_{1 i} $ and $ \xi_j $ ($i,j =x,y,z$) are the components 
of the coordinates of the electrons, 
a summation over the repeated indices is implied,  
and the real and imaginary parts of the complex tensor 
$\Theta_{ij}$ are given by 
\begin{eqnarray} 
Re \left( \Theta_{ij} \right) &=& 
\frac{\cos \left(k_0 R \right)+ k_0 R \sin \left(k_0 R \right)}{R^3}
\left( \delta_{ij} - \frac{3 R_i R_j}{R^2} \right)  
\nonumber \\ 
&& - \frac{ k_0^2 \cos \left(k_0 R \right)}{R}
\left( \delta_{ij} - \frac{R_i R_j}{R^2} \right)  
\nonumber \\ 
Im \left( \Theta_{ij} \right) & = &  
\frac{\sin \left(k_0 R \right) - k_0 R \cos \left(k_0 R \right)}{R^3}  
\left( \delta_{ij} - \frac{3 R_i R_j}{R^2} \right)
\nonumber \\ 
&& 
- \frac{ k_0^2 \sin \left(k_0 R \right)}{R} 
\left( \delta_{ij} - \frac{R_i R_j}{R^2} \right).    
\label{AB_inter_1}  
\end{eqnarray} 
If the distance $R$ is relatively not large 
($k_0 R \ll 1$) it follows from (\ref{AB_inter_1}) that 
$\left| Re \left( \Theta_{ij} \right) \right| \gg \left| Im \left( \Theta_{ij} \right) \right| $ 
and the interaction (\ref{AB_inter}) 
practically reduces to the instantaneous interaction 
of two electric dipoles, proportional to $1/R^3$. 
However, at $k_0 R \stackrel{>}{\sim} 1$ 
the use of the instantaneous and retarded forms  
of the electron-electron interaction leads 
to large differences. % \cite{we-paper}. 
The latter will be studied in detail in a forthcoming 
paper. 
 
We first examine the case of a weak EM field, 
where relatively simple formulas can be obtained 
for the ionization probability $p_A^{ion}$.  
The weak field case is described by the %{\it 1B}  
first order perturbation theory with respect to 
the interaction $\hat{W}$ that is valid when 
$\max\{ W_{e,0}^B, \Gamma_i \} \, \,  T \ll 1$, 
where %$W_{e,0}^B = \left\langle \chi_e \left|\hat{W}_0^+\right| \chi_0 \right\rangle $, 
$ W_{e,0}^B = < \chi_e | \hat{W}_0^+ | \chi_0 > $,   
%$\Gamma_i = 2 \pi \int d\Omega_{\bf p} p_0 \left| W_{{\bf p}_0,0}^A \right|^2$ 
$\Gamma_i = 2 \pi \int d\Omega_{\bf p} p_0 | W_{{\bf p}_0,0}^A |^2$ 
($ W_{{\bf p},0}^A =  < u_{\bf p} | \hat{W}_0^+ | u_0 > $)
is the width of the ground state of the atom $A$ caused by its (direct) photo decay 
and $T$ is the pulse duration of the EM field. 
Assuming for definiteness that the field is instantaneously switched on at $t=0$, 
when both atoms are in the ground states,    
we obtain that the probability to find the atom $A$ in its continuum states   
is, for sufficiently long pulses 
($\overline{\Gamma}_a T > 1$), given by  
\begin{eqnarray} 
p_A^{ion}(T)= \int d\Omega_{\bf p} \left| W_{{\bf p}_0,0}^A + 
\frac{V_{{\bf p}_0,a}^{ee} \, \, \tilde{W}_{e,0}^B}%
{\epsilon_0+\omega_0-\epsilon_e + i \overline{\Gamma}_a/2}\right|^2 T.   
\label{weak-field-lim-1}
\end{eqnarray}
Here, $\Omega_{\bf p}$ and $|{\bf p}_0| = \sqrt{2(\varepsilon_0+\omega_0)}$ are 
the solid angle and the absolute value of the momentum of 
the emitted electron, respectively,    
%$W_{{\bf p},0}^A =  \left\langle u_{\bf p} \left|\hat{W}_0^+\right|u_0 \right\rangle$,  
%$W_{e,0}^B = \left\langle \chi_e \left|\hat{W}_0^+\right| \chi_0 \right\rangle $ 
$V_{{\bf p},a}^{ee} = <  u_{\bf p},\chi_0 | \hat{V}^{ee} | u_0,\chi_e > $,  
$\overline{\Gamma}_a = \Gamma_a + \Gamma_r$ is the total 
width of the state $\psi_a$, where $\Gamma_a$ and $\Gamma_r$ are the contributions 
caused by two-center autoionization and spontaneous radiative decay 
of the excited state of the atom $B$, respectively, and 
\begin{eqnarray} 
\tilde{W}_{e,0}^B = W_{e,0}^B + \int d^3 {\bf p} 
\frac{V_{a,{\bf p}}^{ee}\,\,W_{{\bf p},0}^A}{\varepsilon_0+\omega_0-\varepsilon_p+i0}.   
\label{weak-field-lim-2}
\end{eqnarray} 
%As one could expect, in the case of long pulses 
%the first order perturbation theory predicts  
%a linear growth in the ionization probability with $T$ \cite{ff}. 

Eqs.(\ref{weak-field-lim-1})-(\ref{weak-field-lim-2}) 
show that there are three qualitatively different 
pathways for ionization of the atom $A$.  
(i) The atom $A$ is directly ionized by the EM  
field %(the transitions $u_0 \to u_{{\bf p}_0} $)   
without any participation of the atom $B$. 
(ii) The field excites the atom $B$ into 
the state $\chi_e$;  % (the transition $\chi_0 \to \chi_e$);  
the latter subsequently deexcites 
by transferring the excess of energy to the electron 
of atom $A$ which leads to its ionization. 
(iii) The EM field drives the electron of atom 
$A$ into the continuum % ($u_0 \to u_{{\bf p}} $) 
but the electron returns back into the ground state $u_0$ due 
to the two-center electron-electron 
interaction and only afterwards the same interaction 
transfers the electron into the final 
continuum state $u_{{\bf p}_0}$.  
The pathways (ii) and (iii) are resonant and become efficient only if 
the frequency $\omega_0$ lies in the interval 
$ \Delta E_B - \overline{\Gamma}_a \stackrel{<}{\sim} % 
\omega_0 \stackrel{<}{\sim} \Delta E_B + \overline{\Gamma}_a  $. 
%$ \epsilon_e - \epsilon_0 - \overline{\Gamma}_a \stackrel{<}{\sim} % 
%\omega_0 \stackrel{<}{\sim} \epsilon_e - \epsilon_0 + \overline{\Gamma}_a  $. 

Assuming that $k_0 R \ll 1$, the partial contribution of 
the pathway (ii) (2CPI) reads      
\begin{eqnarray} 
p_{\rm 2CPI} &=& \frac{ T }{2 \pi} % 
\frac{ \Gamma_a \, \, |W_{e,0}^B|^2}% 
{\left( \epsilon_0+\omega_0-\epsilon_e\right)^2 + i \overline{\Gamma}_a^2/4}.  
\label{2CPE}
\end{eqnarray}
It can be compared with 
the probability for photoionization 
of an isolated atom $A$ given by 
$p_{\rm direct}=\frac{ T }{2 \pi} \Gamma_i$.   
At distances $R$, where $\Gamma_a < \Gamma_r$, 
the ratio for the integral contributions  % strengths 
of these two channels is given by 
\begin{eqnarray} 
\eta \sim \frac{ p_{\rm 2CPI} \times \overline{\Gamma}_a }{p_{\rm direct} \times \Delta \omega_0 }
\approx 
\left(\frac{ c/\omega_0 }{ R }\right)^3 
\left(\frac{ a_0}{ R }\right)^3
\frac{ 1/a_0  }{ Z_B^2 \Delta \omega_0 }.   
\label{ratio-integral}
\end{eqnarray}
where $a_0$ is the Bohr radius and $ \Delta \omega_0 $ 
is the spectral width of the EM field. 
When $R$ decreases, $\eta$ becomes less steeply 
dependent on $R$ and eventually $R$-independent 
at distances, where $\Gamma_a > \Gamma_r$.     
Although 2CPI is effective only in the vicinity of the resonance 
while the direct channel may act for 
the whole width $ \Delta \omega_0 \gg \overline{\Gamma}_a $, 
the ratio $\eta$ can be quite large. 

Indeed, let us consider, for instance, a van-der-Waals hetero-dimer such as NaKr (or LiAr) \cite{NaKr}  
in the electronic ground state. For $R \approx 10$\AA\,,  
$\omega_0\approx 10$\,eV, corresponding to the $4p$-$5s$ transition in Kr,  
and assuming $\Delta\omega_0 \sim 1$\,meV,  
the PI of Na ($I_A=5.14$\,eV) is enhanced by $\eta \sim 10^4$ due to 2CPI. 
This number may be considered a lower bound of the enhancement effect 
at the real equilibrium distance $R\approx 5$\AA, where the assumption $\Gamma_a < \Gamma_r$ 
underlying Eq.\,(\ref{ratio-integral}) might not hold. 

Another suitable van-der-Waals dimer is $^7$Li$^4$He. 
It is a very weakly bound and largely extended molecule. The mean internuclear separation 
amounts to $R\approx 30$-$40$\AA, according to theoretical calculations \cite{Kleinekathoefer}. 
Remarkably, even at distances that large, 2CPI triples the ionization yield from Li 
at $\omega_0 \approx 21$\,eV (and assuming again that $\Delta\omega_0 \sim 1$\,meV), 
which corresponds to the $1s^2\,^1S$--$1s2p\,^1P$ transition in He. 
(Note besides  
that the retardation effects contribute here at a level of $10$-$15 \%$.) 
An alternative way of observing 2CPI in the Li-He system might employ Li 
atoms attached to helium nanodroplets ($R \approx 5$ \AA). 
Spectroscopic studies of alkali atoms embedded into liquid helium 
have become possible in recent years \cite{liquidHe}. 

Note that 2CPI may also occur in biological tissue after absorption of 
light or UV radiation. In fact, the process resembles the energy
transfer between organic molecules via F\"orster resonances and related 
bystander effects \cite{Forster}. The main difference is that 2CPI
involves a resonant coupling to the continuum.

\begin{figure}[t]  
\vspace{-0.25cm}
\begin{center}
\includegraphics[width=0.31\textwidth]{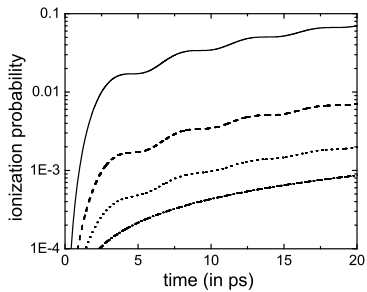}
\end{center}
\vspace{-0.75cm} 
\caption{ \footnotesize{ 
Ionization probability for Li in the presence of He as a function of time.  
$F_0=10^{-4}$ a.u., the field frequency is resonant to the 
$1s^2\,^1S$--$1s2p\,^1P$ % 1s$^2$ $^1S$ $\to$ $1s 2p$ $^1P$ 
transition in He. % ($\omega_0 \approx 21$ eV).  
The solid, dash and dot curves display results 
for $R=5$, $7.5$ and $10$ \AA, respectively.   
For comparison, the ionization probability 
for an isolated Li atom 
in the same EM field is shown by the dash-dot curve. } } 
\label{figure2}
\end{figure}

Let us now turn to ionization in stronger EM fields,  
when $\max\{ W_{e,0}^B, \Gamma_i \} T \stackrel{>}{\sim} 1$, 
which can be considered using the rotating wave approximation. % {\it RWA}.  
Compared to the weak-field case, ionization 
of the atom $A$ due to the presence of the atom $B$ is now 
enhanced less dramatically, but acquires interesting new features. 
In figure \ref{figure2} we show the probability of ionization 
of a Li atom in the electromagnetic field with 
$F_0 = 10^{-4}$ a.u. ($I = 3.5 \times 10^{8}$ W/cm$^2$) 
when a He atom is located  nearby 
(at $ R=5 $, $7.5$ and $10$ \AA). 
The frequency $\omega_0 \approx 21.$ eV is 
chosen to be resonant to 
the $1s^2\,^1S$--$1s2p\,^1P$ % $1s^2 1 ^1 S$ $\to$ $1s 2p  ^1P$ 
transition in $He$. For simplicity ${\bf F}_0$ 
is assumed to be directed along ${\bf R}$ 
\cite{explanation}. 
For comparison the ionization probability of an isolated 
Li atom is also displayed. 

%For the parameters chosen one has $\Gamma_i = 10^{-9} $ a.u., 
%$|W_{e,0}^B|= 1.6 \times 10^{-5}$ a.u. 
%$\Gamma_r = 1.5 \times 10^{-8}$ a.u. and 
%$\Gamma_a = 1.8 \times 10^{-7}$, $1.56 \times 10^{-8}$ and 
%$2.8 \times 10^{-9}$ a.u. for $R=5$, $7.5$ and $10$ \AA, respectively.   
%It is seen in the figure that 
%compared to the ionization probability of an isolated 
%Li atom, which simply monotonously increases with time,  
%ionization in the presence of a He atom 
%shows a step-wise behavior in which time intervals, 
%where the ionization probability rapidly increases, 
%are separated by intervals, where the probability 
%remains almost a constant. 
 
For the parameters chosen one has $\Gamma_i = 2.7 \times 10^{-8} $ eV, 
$|W_{e,0}^B|= 4.3 \times 10^{-4}$ eV,  
$\Gamma_r = 4 \times 10^{-7}$ eV and 
$\Gamma_a = 4.9 \times 10^{-6}$, $4.2 \times 10^{-7}$ and 
$7.6 \times 10^{-8}$ eV for $R=5$, $7.5$ and $10$ \AA, respectively.   
It is seen in the figure that 
compared to the ionization probability of an isolated 
Li atom, which simply monotonously increases with time,  
ionization in the presence of a He atom 
shows a step-wise behavior in which time intervals, 
where the ionization probability rapidly increases, 
are separated by intervals, where the probability 
remains almost constant. 

The origin of this is rather simple.    
In the resonant EM field the population of He oscillates  
between its ground and excited states with the frequency $\Omega = 2 |W_{e,0}^B| $.  
Once the population of the excited state becomes noticeable,  
the two-center autoionization comes into play 
opening the additional pathway for ionization of the Li atom. 
When most of the population has returned back into the ground state 
of He the two-center autoionization effectively switches off 
and the ionization process weakens. 
Since $\Gamma_a$ exceeds $\Gamma_i$ 
for all the three distances,  
for all of them the autoionization channel 
has a strong overall effect on the ionization of Li.  

\begin{figure}[t]  
\vspace{-0.25cm}
\begin{center}
\includegraphics[width=0.42\textwidth]{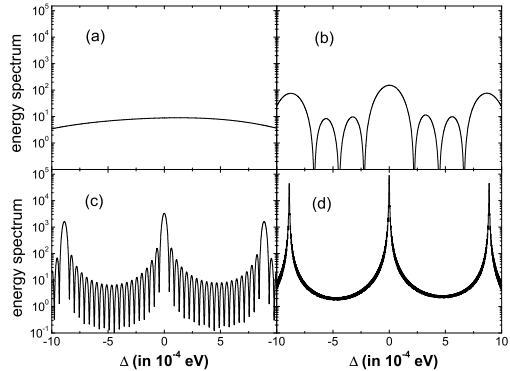}
\end{center}
\vspace{-0.75cm} 
\caption{ \footnotesize{ 
Energy spectrum of the emitted electrons, as a function 
of $\Delta = \varepsilon_p - \varepsilon_0 - \omega_0$,  
for the same atomic system and field as in fig. \ref{figure2},  
but only for $R=5$ \AA. Panels (a), (b), (c) and (d) show the spectrum for 
the field pulse duration  
%$T = 3.7 \times 10^{-12}$, $1.9 \times 10^{-11}$, 
%$9.2 \times 10^{-11}$ and $9.2 \times 10^{-10}$ s, 
$T = 3.7 $, $19$, $92$ and $920 $ ps,
respectively. } } 
\label{figure3}
\end{figure}

Additional insight into the ionization process is 
obtained by considering the energy spectrum 
of emitted electrons. 
Such a spectrum is shown in figure \ref{figure3} 
for four different values of $T$ that 
enables one to trace the formation of the spectrum in time. 
The atomic system and the EM field are the same as in figure \ref{figure2} 
and $ R=5 $ \AA. In panel (a) the pulse duration 
is so short that the spectrum does not yet possess any substantial structure. 
However, in panel (b) one can already see in the spectrum three main maxima  
which develop into very pronounced peaks 
with a further increase in $T$ (see panels (c) and (d)). 
The origin of these peaks is similar to the splitting  
into three lines of the energy spectrum of photons 
emitted during atomic fluorescence in a resonant EM field \cite{mollow1969}.  
In such a field the ground and excited levels of $He$ split into 
two sub-levels, which in our case differ by $\Omega $:   
$\epsilon_0$ $\to$ $\epsilon_0 \pm \Omega/2$ and 
$\epsilon_e$ $\to$ $\epsilon_e \pm \Omega/2$. 
As a result, when undergoing autoionizing transitions the energy transfer 
to Li peaks at $\omega_0$ and  $\omega_0 \pm \Omega/2$. 
Additional multiple maxima, seen in fig. \ref{figure3} when the condition 
$\Gamma_a T \ll 1$ is fulfilled, are related to the finiteness of the pulse 
duration and the distance between them is roughly given by $2 \pi/T$.  

In conclusion, photoionization of an atomic center $A$ can 
change dramatically in the presence of a neighboring 
center $B$ at nanometer distances provided one of 
the transition frequencies of the latter is close to the field frequency. 
This resonance effect is especially strong in 
the case of weak EM fields when it may enhance 
photoionization by orders of magnitude. 
In stronger EM fields photoionization 
acquires new interesting features. In particular, 
a step-wise increase in the ionization probability 
with time and a splitting of the photoelectron spectrum 
into three prominent lines, similar to resonance fluorescence, arise.    
This efficient two-center ionization mechanism may also play 
a significant role in chemical and biological systems. %, rendering it of  where it can substantially affect 
%the dynamics and time evolution of the system. 


\begin{thebibliography}{99}
\bibitem{at-units} Atomic units (a.u.) are used unless otherwise stated.
\end{thebibliography}
\end{document}